\documentclass[10pt,letterpaper]{article}
\usepackage[top=0.85in,left=2.75in,footskip=0.75in]{geometry}

\usepackage{amsmath,amssymb}

\usepackage{changepage}

\usepackage[utf8x]{inputenc}

\usepackage{textcomp,marvosym}

\usepackage{cite}

\usepackage{nameref,hyperref}

\usepackage[right]{lineno}

\usepackage{microtype}
\DisableLigatures[f]{encoding = *, family = * }

\usepackage[table]{xcolor}

\usepackage{array}

\usepackage[FIGTOPCAP, raggedright, nooneline]{subfigure}
\usepackage[separate-uncertainty,range-phrase = --]{siunitx}
\usepackage{float}
\newcolumntype{+}{!{\vrule width 2pt}}

\newlength\savedwidth

\newcommand\thickhline{\noalign{\global\savedwidth\arrayrulewidth\global\arrayrulewidth 2pt}%
\hline
\noalign{\global\arrayrulewidth\savedwidth}}

\raggedright
\usepackage[aboveskip=1pt,labelfont=bf,labelsep=period,justification=raggedright,singlelinecheck=off]{caption}

\makeatletter
\renewcommand{\@biblabel}[1]{\quad#1.}
\makeatother

\usepackage{lastpage,fancyhdr,graphicx}
\usepackage{epstopdf}
\fancyheadoffset[L]{2.25in}
\fancyfootoffset[L]{2.25in}

\begin{document}
\vspace*{0.2in}

\begin{flushleft}
{\Large
\textbf\newline{Tissue evolution: Mechanical interplay of adhesion, pressure, and heterogeneity} 
}
\newline
\\
Tobias Büscher\textsuperscript{1},
Nirmalendu Ganai\textsuperscript{1,2},
Gerhard Gompper\textsuperscript{1},
Jens Elgeti\textsuperscript{1*},
\\
\bigskip
\textbf{1} Theoretical Soft Matter and Biophysics, Institute of
Complex Systems and Institute for Advanced Simulation, Forschungszentrum Jülich, 52425 Jülich, Germany
\\
\textbf{2} Department of Physics, Nabadwip Vidyasagar College, Nabadwip, Nadia 741302, India
\\
\bigskip

%
%





* j.elgeti@fz-juelich.de

\end{flushleft}
\section*{Abstract}
The evolution of various competing cell types in tissues, and the
resulting persistent tissue population, is studied numerically and analytically
in a particle-based model of active tissues.
Mutations change the properties of cells in various ways, including
their mechanical properties. Each mutation results in an advantage or disadvantage
to grow in the competition between different cell types. While changes in signaling
processes and biochemistry play an important role, we focus on changes in the mechanical properties by studying the result
of variation of growth force and adhesive cross-interactions between cell types.
For independent mutations of growth force and adhesion strength, the tissue evolves
towards cell types with high growth force and low internal adhesion strength, as both increase the homeostatic pressure.
Motivated by biological evidence, we postulate a coupling between both parameters, such that an increased growth force comes at the cost of a higher internal adhesion strength or vice versa. 
This tradeoff controls the evolution of the tissue, ranging from unidirectional evolution to very heterogeneous and dynamic populations. 
The special case of two competing cell types reveals three distinct parameter regimes: Two in which one cell type outcompetes the other,       
and one in which both cell types coexist in a highly mixed state. Interestingly, a single
mutated cell alone suffices to reach the mixed state, while
a finite mutation rate affects the results only weakly. 
Finally, the coupling between changes in growth force and adhesion strength
reveals a mechanical explanation for the evolution towards intra-tumor heterogeneity,
in which multiple species coexist even under a constant evolutianary pressure.

\section*{Introduction}
Mutations change the cell fitness and thus its chance to survive and proliferate \cite{weinbergbook}.
Advantageous mutations are more likely to persist due to natural selection, which drives the evolution of a tissue towards fitter cells \cite{Greaves2012}. Cancer represents an example of evolution on a short time scale \cite{Bozic2013}. Furthermore, cancer is a multistep process, i.e. several mutations are needed for a tumor in order to develop and become malignant \cite{Vogelstein1993}. Hence, tumorigenesis might be expected to happen in a serial manner, i.e. a cell acquiring a "beneficial" mutation and taking over the whole tissue. After some time, a daughter cell acquires another mutation and again takes over. 
Interestingly, however, tumors do not consist of a single cell type, but instead several subpopulations coexist within the same tumor. This is called intra-tumor heterogeneity \cite{Heppner1984}. 

Each mutation changes certain biochemical properties of a cell. This ranges from misfunction in the error correction machinery during DNA replication and disruptions in signaling pathways  to epigenetic changes in the expression level of certain proteins \cite{weinbergbook, Preston2010, Schnekenburger2012}. 
All  these changes can also affect the mechanical properties of the mutated cell, e.g. mutated cells which express less adhesion proteins might be able to detach from the primary tumor more easily \cite{Petrova2016}, necessary to form metastases. On the other hand, mechanics feeds back onto growth in several ways, e.g. increased apoptosis rate due to mechanical stresses \cite{Wernig2002, Cheng2009} or dependence of the growth of tissue spheroids on the properties of the surrounding medium \cite{Montel2011, Alessandri2013, Helmlinger1997}. 

It is the mechanical contribution to tissue development that we want to focus on in this work.
For mechanically regulated growth, homeostatic pressure plays an important role \cite{Basan2009}.
In the homeostatic state, when apoptosis and division balance each other, a tissue exerts a certain pressure onto its surrounding, the homeostatic pressure $P_{\text{H}}$. The tissue is able to grow as long as the external pressure $P$ is smaller than $P_{\text{H}}$. 
For the competition between different tissues for space, it has been suggested that the tissue with the higher homeostatic pressure grows at the expense of the weaker tissue. 
Several theoretical studies employ this concept in order to describe interface propagation between two competing tissues \cite{Williamson2018, Ranft2014, Podewitz2016}.
A metastasis would need to reach a critical size, below which the additional Laplace pressure due to surface tension would cause the metastasis to shrink and disappear \cite{Basan2009}. 
However, reduced adhesion between tissues, which increases surface tension, leads to an enhanced growth rate at the interface between them, stabilizing coexistence even for differing homeostatic pressures \cite{Ganai2019}.

In this work, we study the influence of mutations that change the mechanical properties of cells on the competition dynamics, especially the interplay between changes in the adhesive properties and the strength with which a cell pushes onto its surrounding. Particularly interesting is the case where loss of adhesion comes at the cost of lower growth strength. This is motivated by the observed down-regulation of E-cadherin, an adhesion protein in epithelia, in many types of cancer \cite{Beavon2000}. Interestingly, E-cadherin is also involved in signaling processes connected to cell growth \cite{Pece2008}. We find that in this case several cell types with different mechanical properties can coexist and that the cell type with the highest homeostatic pressure does not necessarily dominate the competition.

\section*{Results}
Several models have been developed previously in order to study tissue growth \cite{VanLiedekerke2015}, in combination with different simulation techniques, including vertex \cite{Farhadifar2007, Alt2017} and particle-based \cite{Drasdo2005, Schaller2005} models as well as Cellular Potts models \cite{Graner1992, Szabo2013}. 
We employ the two particle growth (2PG) model of Refs. \cite{Basan2011a, Podewitz2015, Ganai2019}. A cell is described by two particles which repel each other via a growth force
\begin{equation}
\textbf{\textit{F}}_{ij}^{\text{G}} =
\frac{G}{(r_{ij} + r_0)^2}\hat{\textbf{\textit{r}}}_{ij}\text{,}\label{growth_force}
\end{equation}
with strength $G$, unit vector $\hat{\textbf{\textit{r}}}_{ij}$, distance $r_{ij}$ between the two particles and a constant $r_0$.
Different cells interact via a soft repulsive force $\textbf{\textit{F}}_{ij}^{\text{V}}$ on short distances, maintaining an excluded volume, and a constant attractive force  $\textbf{\textit{F}}_{ij}^{\text{A}}$ on intermediate distances, modeling cell-cell adhesion, with
\begin{equation}
\left.
\begin{array}{@{}ll@{}}
\textbf{\textit{F}}_{ij}^{\text{V}} &= f_0\left(\frac{R_{\text{PP}}^5}{r_{ij}^5}-1\right)\hat{\textbf{\textit{r}}}_{ij} \label{volume}\\
\textbf{\textit{F}}_{ij}^{\text{A}}& = -f_1\hat{\textbf{\textit{r}}}_{ij} 

\end{array}\right\} \text{for } r_{ij}<R_{\text{PP}} \text{,}
\end{equation}
with exclusion coefficient $f_0$, adhesion strength coefficient $f_1$, and cut-off length $R_{\text{PP}}$. 
A cell divides when the distance between its two particles reaches a size threshold $r_{\text{ct}}$. A new particle is then placed close (randomly within a short distance $r_{\text{d}}$) to each of the two particles of the divided cell. Each of these pairs then constitutes a new cell. Apoptosis is modeled by removing cells randomly at a constant rate $k_{\text{a}}$. 

We employ a dissipative particle dynamics-type thermostat, with an effective temperature $T$, to account for energy dissipation and random fluctuations. 
We choose the value of $T$ such that cells can escape local minima, but other thermal effects are negligible.
Note that all parameters can be set individually for each cell type as well as between different cell types for inter-cell interactions. We only vary the growth-force strength $G^{\alpha}$ and adhesion strength $f_1^{\alpha\beta}$ between cells of the same ($\alpha=\beta$) and different ($\alpha\neq\beta$) cell types, respectively, where $\alpha$ and $\beta$ are cell-type numbers. We report simulation parameters relative to a standard host cell type (see Materials and methods for numerical values), denoted with a dagger, e.g. $G^\dagger = G/G^0$. Time is measured in terms of the inverse apoptosis rate $k_{\text{a}}$, distance in units of the pair potential cut-off length $R_{\text{PP}}$ and stresses in units of $G^0/R_{\text{PP}}^{4}$. Quantities reported in these units are denoted by an asterisk~$^*$. All simulations are performed in a cubic box with edge length $L=12\cdot R_{\text{PP}}$ and periodic boundary conditions in all directions, unless stated otherwise.

\begin{figure}
	\begin{adjustwidth}{-2.25in}{0in}
	\centering
	\includegraphics[width=1.\linewidth]{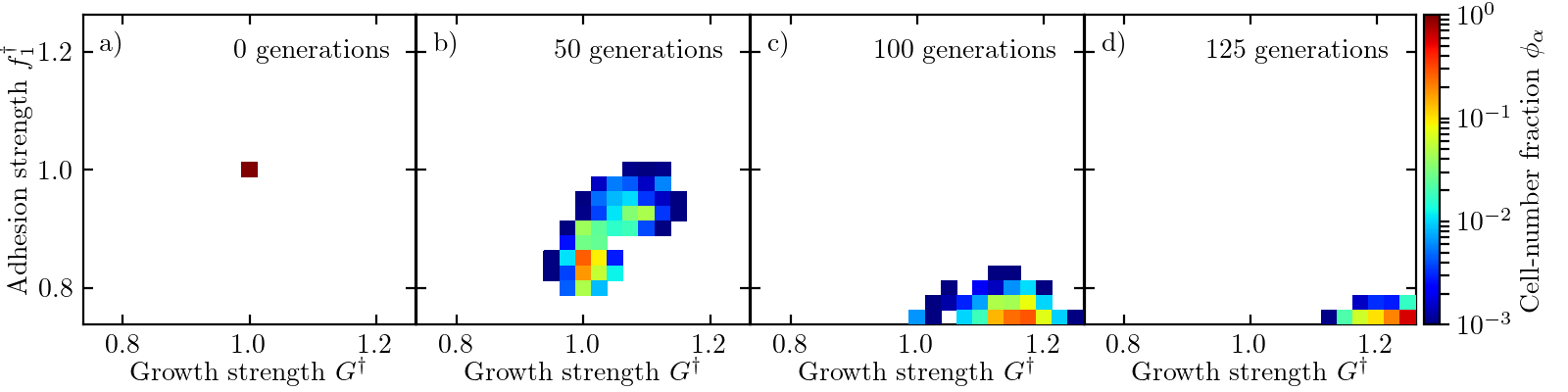}

	\caption{Evolution of a tissue with mutations altering growth-force strengh $G^{\dagger}$ and adhesion strength $f_1^{\dagger}$ independently. Heatmaps displaying cell-number fractions $\phi_\alpha$ after a)~zero generations (initial condition), b)~50 generations, c)~100 generations and d)~125 generations.}
	\label{free_evolution}
\end{adjustwidth}
\end{figure} 
Tumor cells even within the same tumor are not all identical, but vary in terms of all kind of attributes, e.g. expression levels of different proteins \cite{Marusyk2012} or their reaction to certain treatments \cite{Marusyk2010}. Hence, there is not only a competition between the tumor and the host, but also between cell-subpopulations of the tumor.  Different models exist to describe tumor heterogeneity, e.g. cancer stem cells \cite{Shackleton2009} or clonal evolution \cite{Nowell1976}. In the latter case, a tumor originates from a single mutated cell, which can acquire additional mutations over time, yielding additional subpopulations.  We model this behaviour by defining a fixed number $n$ of different "genotypes", each having a different growth-force strength $G^\alpha$ and adhesion strength $f_1^{\alpha\alpha}$.  Mutations are implemented by offering each daughter cell after a division event the chance to change its genotype with a certain probability. 

In tissues, several adhesion mechanisms exist, serving a variety of different functions to maintain tissue integrity. Between epithelial cells, the strength of cell-cell adhesion is to a large degree regulated by anchoring junctions, e.g. adherens junctions, which connect the actin cytosceletons of neighbouring cells. Adherens junctions are mediated by cadherins, which form homophilic bonds between cells. Thus, the strength of adhesion between cells is limited by the cell expressing less cadherin, or, in terms of our simulation model $f_{1}^{\alpha\beta} = \min(f_{1}^{\alpha\alpha}, f_{1}^{\beta\beta})$.  A reduced adhesion strength yields a higher homeostatic pressure \cite{Podewitz2015}, which is otherwise dominated by the growth-force strength $G$. For free parameter evolution, the tissue thus evolves to a strong-growing and low-adhesive genotype (see Fig~\ref{free_evolution}), as predicted by the homeostatic pressure approach \cite{Basan2009}.
\begin{figure}
	\begin{adjustwidth}{-2.25in}{0in}
		\centering
		\includegraphics[width=1.\linewidth]{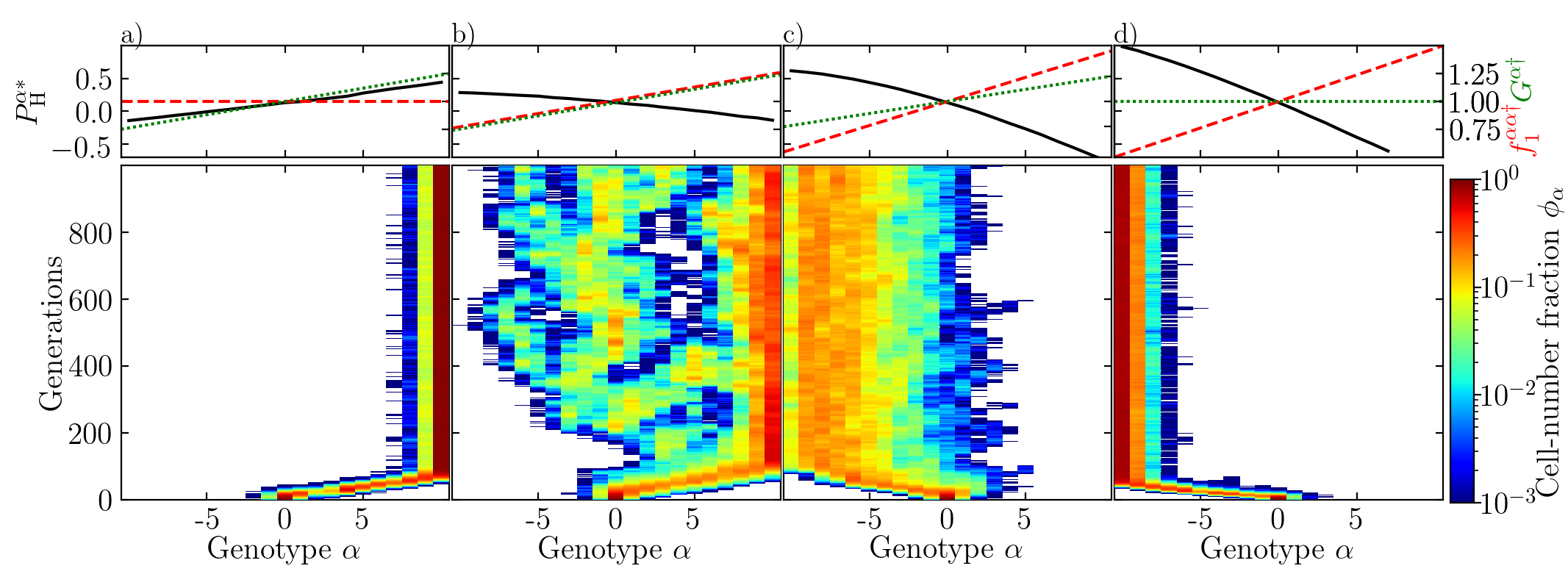}	
		
		\caption{ Time evolution of the cell-number fractions $\phi_{\alpha}$ of each genotype for tradeoff paramter a) $\tau=0$, b) $\tau=1$, c) $\tau=2$ and d) $\tau\rightarrow\infty$, $d\rightarrow 0$. Simulations start from a host (standard) tissue at homeostasis, with $n=21$ genotypes, $p_\text{m}=0.01$ in all and $d=0.025$ in a)-c). White space corresponds to times where no cells of the genotype exist. Color is coded on a logarithmic scale. Curves above display homeostatic pressure $P_{\text{H}}^{\alpha*}$ (black solid), growth-force strength $G^{\alpha\dagger}$(red dashed) and self adhesion strength $f_1^{\alpha\alpha\dagger}$ (green dotted) of the corresponding genotype.}
		\label{heterogeneity}
	\end{adjustwidth}
\end{figure}

However, E-cadherin also plays a role in signaling processes connected to cell growth, and thus a reduced expression might come at the cost of a lower growth-force strength $G$, which in turn yields a lower homeostatic pressure. 
We thus turn our attention to the case where an increase in growth-force strength $G^\alpha$ comes at the cost of a higher self-adhesion strength $f_1^{\alpha\alpha}$. We assume the relations as

\begin{align}
G^\alpha &= (1+D^\alpha )G^0 \\
f_1^{\alpha\alpha}&= (1+D^\alpha\cdot \tau)f_1^0 \text{,} \label{genotype_values}
\end{align}
with genotype number $\alpha$ in the range $[-(n-1)/2, (n-1)/2]$, evolutionary distance $D^\alpha=d\cdot\alpha$, distance $d$ between neighbouring genotypes
and tradeoff paramteter~$\tau$ (with $G^\alpha, f_1^{\alpha\alpha}>0\ \forall\ \alpha$). After a division event, each daughter cell might mutate into a new genotype with probability $p_{\text{m}}$. If the cell mutates, its genotype number is changed to $\alpha_{\text{mother}}~\pm1$ randomly. This yields a mutation rate $k_{\text{m}}=2p_{\text{m}}k_{\text{a}}$.

\begin{figure}
	\begin{adjustwidth}{-2.25in}{0in}
		\flushleft{\hspace{.2cm}a)\hspace{6.cm}b)\hspace{6cm}c)}\\
		\includegraphics[width=0.32\linewidth]{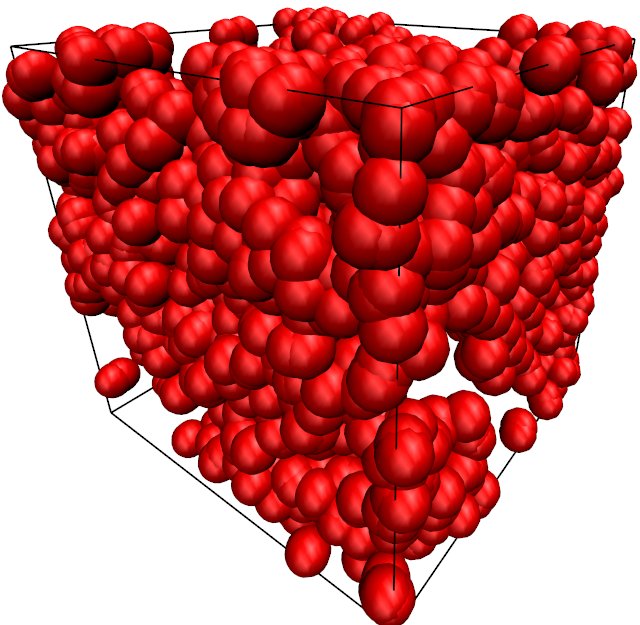}
		\includegraphics[width=0.32\linewidth]{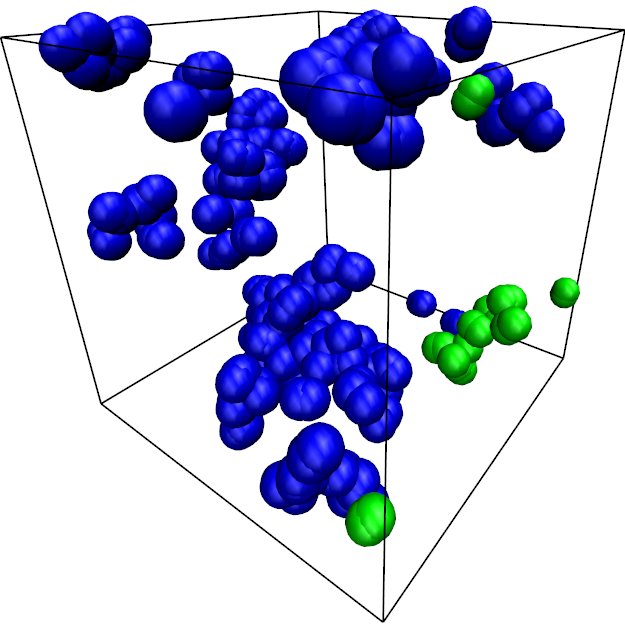}
		\includegraphics[width=0.32\linewidth]{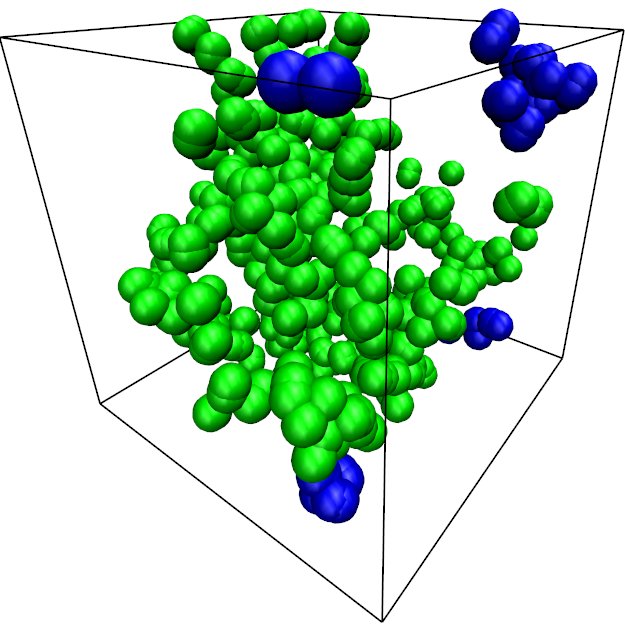}
		\caption{Simulation snapshots obtained from the simulation shown in Fig~\ref{heterogeneity}b). a)~The dominating genotype ($\phi_{10} = 0.283$) after 1000 generations. b)~Genotype $\alpha=-6$ (blue) and $\alpha=6$ (green) after 485 generations ($\phi_{-6} = 0.027$, $\phi_{6} = 0.002$). c)~Same as b), but after 910 generations ($\phi_{-6} = 0.003$, $\phi_{6} = 0.045$).}
		\label{snapshot}
	\end{adjustwidth}
\end{figure}

Figure \ref{heterogeneity} displays results of such simulations for four different cases: only variation of growth-force strength ($\tau=0$), balanced tradeoff ($\tau=1$), adhesion strength varied twice as much as growth-force strength ($\tau=2$) and only variation of adhesion strength ($\tau\rightarrow\infty$). Without tradeoff (Fig~\ref{heterogeneity}a)), the tissue evolves towards the strongest growing genotype or, equivalently, the one with the highest homeostatic pressure. Similarly, for $\tau\rightarrow\infty$ (Fig~\ref{heterogeneity}d)), the system evolves towards the lowest adhesive genotype (again, the one with the highest $P_{\text{H}}$). We find the most dynamic evolution for a balanced tradeoff (Fig~\ref{snapshot} and \ref{heterogeneity}b)). At first, the system evolves to stronger growing and more adhesive genotypes. Over time a noticable fraction of cells evolves also towards weak-growing, less adhesive genotypes. The cell-number fractions $\phi_{\alpha}=N_\alpha/N$ (with individual and total number of cells, $N_{\alpha}$ and $N$), show large fluctuations (see Fig~\ref{snapshot}b) and c)), with individual genotypes not being populated at all for certain time periods. Besides this highly dynamic temporal evolution, after an initial time period the system is dominated by genotypes with increased growth force and adhesion strength at all times, with the one at the upper boundary having the highest cell-number fraction for most of the time (see Fig~\ref{snapshot}a)). This result comes at a surprise, as this is also the genotype with the lowest homeostatic pressure, while the one at the lower boundary, which is basically never populated, has the highest $P_{\text{H}}$. For a higher tradeoff (Fig~\ref{heterogeneity}c)), we still find a broad distribution of genotypes, with less adhesive genotypes dominating over the stronger growing ones, i.e. the loss in growth-force strength is overcompensated by a lower adhesion strength.

\begin{figure}
	\begin{adjustwidth}{-2.25in}{0in}

	\flushleft{\hspace{.2cm}(a)\hspace{9.cm}(b)}\\

	\includegraphics[width=0.49\linewidth]{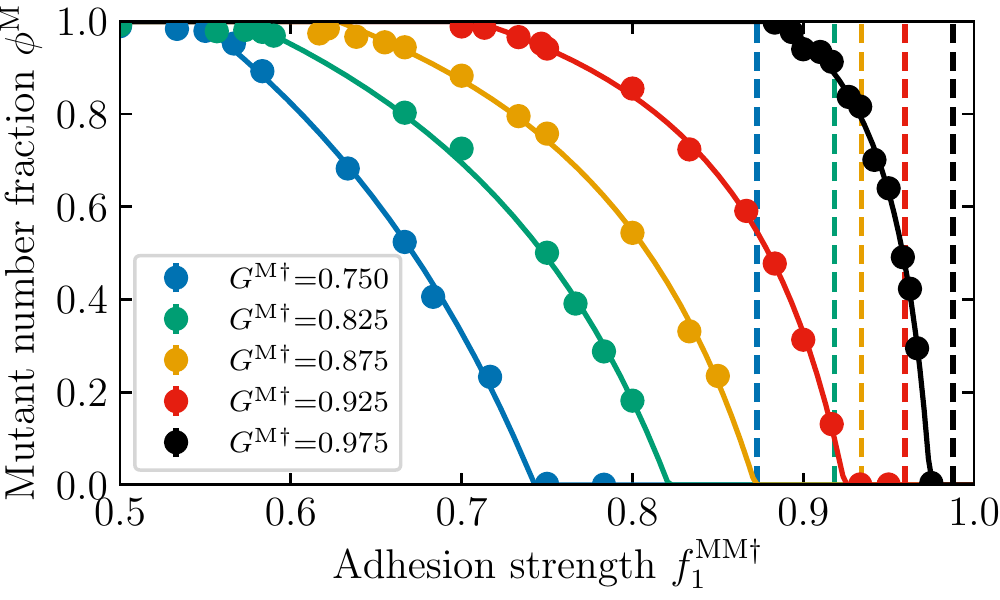}	
	\includegraphics[width=0.49\linewidth]{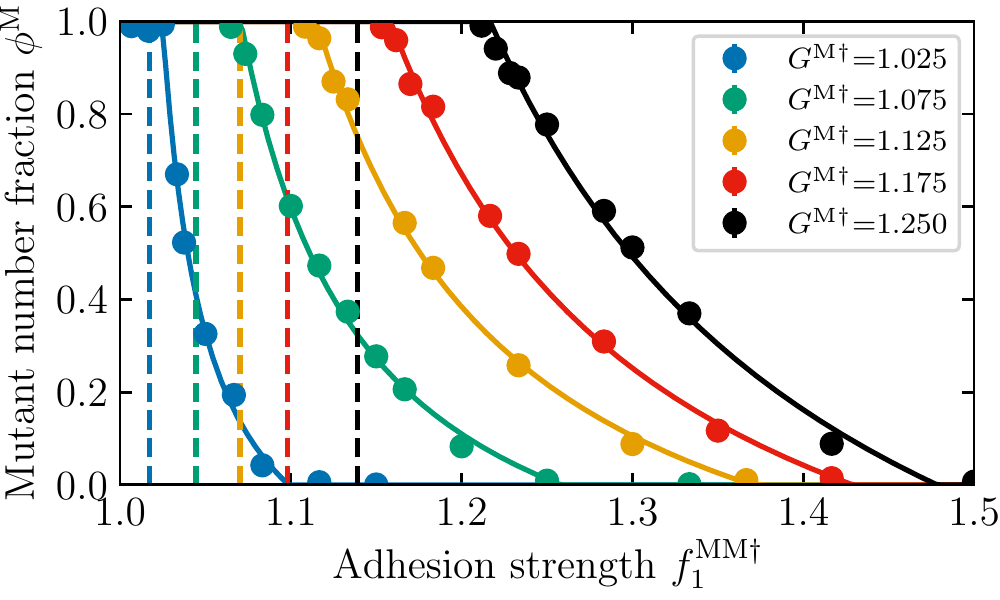}

	\caption{a) Average number fraction $\phi^{\text{M}}$ of the mutant in terms of its adhesion strength $f_1^{\text{MM}\dagger}$ for various (reduced) growth-force strengths $G^{\text{M}\dagger}$. Error bars are obtained via block-averaging method (hidden behind markers) \cite{allantildesley}. Dashed vertical lines indicate the points below which the mutant has a higher homeostatic pressure, solid lines are fits to Eq.~\ref{stable_fixed_point_expansion}. b)~Same as in a) but for increased growth force and adhesion strengths of the mutant.}
	\label{number_fractions}
	\end{adjustwidth}
\end{figure}

In order to gain insight into the underlying mechanism of this dynamic evolution, we study the competition between two genotypes and no mutations ($p_{\text{m}}=0$). Simulations are started from a single mutated cell (with increased/decreased growth force and adhesion strength) in a host tissue at the homeostatic state (we label the mutant with \em M \em and the host (wild type) with \em W\em). Even in this simplified case, we find one parameter regime in which the mutant is not able to grow, one regime with stable coexistence in a highly mixed state and another regime in which the mutant outcompetes the host.
Figure \ref{number_fractions} shows the averaged number fractions of the mutant at the steady state. For reduced growth force and adhesion strength (Fig~\ref{number_fractions}a)), the mutant can only grow against the host if its adhesion strength is reduced below a critical $f_1^{\text{crit}}$. In terms of Eq.~\eqref{genotype_values}, the value of $f_1^{\text{crit}}$ roughly corresponds to a balanced tradeoff ($\tau\approx1$). Already for $f_1^{\text{MM}}>f_1^{\text{crit}}$, the homeostatic pressure of the mutant exceeds the one of the host, i.e. a parameter regime exists in which the mutant is not able to grow, despite of the higher $P_{\text{H}}$. The reverse happens when growth force and adhesion strength are increased. The mutant completely takes over the compartment, although its homeostatic pressure is smaller than that of the host. Again, coexistence is only found when the adhesion strength is increased above $f_1^{\text{crit}}$.  In the coexistence regime, the mutant number fraction scales as $\phi^{\text{M}} \propto 1/(f_1^{\text{MM}}-f_1^{\text{WW}})$. 

Altogether, the competition between two genotypes alone yields the same qualitative results as the more complex multi-genotype case discussed before. Still, the question remains how a genotype with lower homeostatic pressure can outcompete a stronger genotype. The answer can only lie in the adhesion strength $f_{1}^{\text{MW}} = \min(f_{1}^{\text{MM}}, f_{1}^{\text{WW}})$ between mutant and host cells. This choice of cross-adhesion strength breaks symmetry, as the stronger adhering genotype has more free space at the interface, which favors divisions \cite{Ganai2019}.

To address this question, we develop a phenomenological model which incorporates pressure-dependent growth as well as interfacial effects, in order to obtain a qualitative explanation of the simulation results.

We start with the expansion of the bulk growth rate $k_{\text{b}}$ around the homeostatic pressure,

\begin{equation}
k_{\text{b}} = \kappa (P-P_{\text{H}})\text{,} \label{bulk}
\end{equation}
with the pressure response coefficient $\kappa$.
Due to the high degree of mixing, the number fractions $\phi^{\text{M/W}}$ and hence the strengh of interfacial effects vary locally. In a mean-field approximation, we take the interfacial effects to be proportional to $\phi^{\text{M}}(1-\phi^{\text{M}})$, with individual prefactors $\Delta k_{\text{s}}^{\text{M/W}}$ for each genotype. The time evolution is then given by

\begin{align}
\partial_t \phi^{\text{M}}= &\kappa(P_{\text{H}}^{\text{M}} - P)\phi^{\text{M}} + \Delta k_{\text{s}}^{\text{M}}\phi^{\text{M}}(1-\phi^{\text{M}}) \label{A}\\
\partial_t (1 - \phi^{\text{M}} ) =&\kappa(P_{\text{H}}^{\text{M}} + \Delta P_{\text{H}}  - P)(1-\phi) 
+ \Delta k_{\text{s}}^{\text{W}}\phi^{\text{M}}(1-\phi^{\text{M}})\text{,} \label{B}
\end{align}
with the difference in homeostatic pressure $\Delta P_{\text{H}} = P_{\text{H}}^{\text{W}}-P_{\text{H}}^{\text{M}}$.
Addition of Eqs. \eqref{A} and \eqref{B} yields the pressure

\begin{align}
P = P_{\text{H}}^{\text{W}} - \Delta P_{\text{H}}\phi^{\text{M}} + \frac{\Delta k_{\text{s}}^{\text{M}} + \Delta k_{\text{s}}^{\text{W}}}{\kappa}\phi^{\text{M}}(1-\phi^{\text{M}})\text{.}\label{pressure}
\end{align}
Thus, the pressure is given by the homeostatic pressures of the two genotypes weighted by their number fraction plus an interfacial term. A figure displaying the pressure measured during the simulations shown in Fig~\ref{number_fractions} can be found in the \nameref{S1_Appendix}.
Insertion of Eq.~\eqref{pressure} into Eq.~\eqref{A} yields a differential equation for the number fraction with three fixed points ($\partial_t\phi^{\text{M}} = 0$),  $\phi_1^{\text{M}}=0$, $\phi_2^{\text{M}}=1$, and 

\begin{align}
\phi_3^{\text{M}} = \frac{-\kappa \Delta P_{\text{H}} + \Delta k_{\text{s}}^{\text{M}}}{\Delta k_{\text{s}}^{\text{M}} + \Delta k_{\text{s}}^{\text{W}}}\text{.}
\label{stable_fixed_point}
\end{align}

We discuss this result for the case of reduced growth force and adhesion strength of the mutant.  $\Delta k_{\text{s}}^{\text{M}}$ might be expected to vanish, as $f_1^{\text{MM}} =f_1^{\text{MW}} $ and mutant cells thus would not feel whether neighbouring cells are mutant or host cells. However, in order to grow, a cell needs to impose a strain on its surrounding. Host cells adhere more strongly to each other, thus it is harder for a mutant cell to impose a strain when surrounded by host cells. Hence,  $\Delta k_{\text{s}}^{\text{M}}$ is actually negative and the homeostatic pressure of the mutant needs to exceed the host pressure by $-\Delta k_{\text{s}}^{\text{M}}/\kappa$ in order to be able to grow against the host. At this point, $\phi_3^{\text{M}}$ becomes positive, as long as $\Delta k_{\text{s}}^{\text{M}} + \Delta k_{\text{s}}^{\text{W}}> 0$. Host cells can impose a strain more easily when surrounded by mutant cells and, additionally, have more free space than when surrounded by other host cells. Hence, $|\Delta k_{\text{s}}^{\text{M}}|< \Delta k_{\text{s}}^{\text{W}}$ and the above mentioned condition is fulfilled.  Similarly, coexistence can be found for increased growth force and adhesion strength when $ \Delta P_{\text{H}}> -\Delta k_{\text{s}}^{\text{W}}/\kappa$. 
The above mentioned scaling of the mutant number fraction can be obtained by an expansion of $\Delta P_{\text{H}}$ and $\Delta k_{\text{s}}^{\text{M/W}}$ to linear order in terms of $\epsilon := (f_1^{\text{MM}}-f_1^{\text{WW}})/f_1^{\text{WW}}$ in Eq.~\eqref{stable_fixed_point},

\begin{align}
\phi_3^{\text{M}} = \frac{-\kappa \Delta P_{\text{H}}^0}{(\Delta k_{\text{s}}^{\text{M1}} + \Delta k_{\text{s}}^{\text{W1}})\epsilon}
+\frac{-\kappa \Delta P_{\text{H}}^1 + \Delta k_{\text{s}}^{\text{M1}}}{\Delta k_{\text{s}}^{\text{M1}} + \Delta k_{\text{s}}^{\text{W1}}}\text{.}
\label{stable_fixed_point_expansion}
\end{align}
The zeroth order terms of $\Delta k_{\text{s}}^{\text{M/W}}$ vanish as there are no interfacial effects when the adhesion strength between host and mutant cells is equal to their self-adhesion strength, while $\Delta P_{\text{H}}^0$ can be non-zero due to a changed growth-force strength. 
Indeed, Eq.~\eqref{stable_fixed_point_expansion} reproduces the simulation data reasonably well (see Fig~\ref{number_fractions}). 
A discussion of the numerical values of the fitted parameters and additional results can be found in \nameref{S1_Appendix}.

\begin{figure}
	\begin{adjustwidth}{-2.25in}{0in}
		\flushleft{\hspace{.2cm}(a)\hspace{9.cm}(b)}\\
		\includegraphics[width=0.49\linewidth]{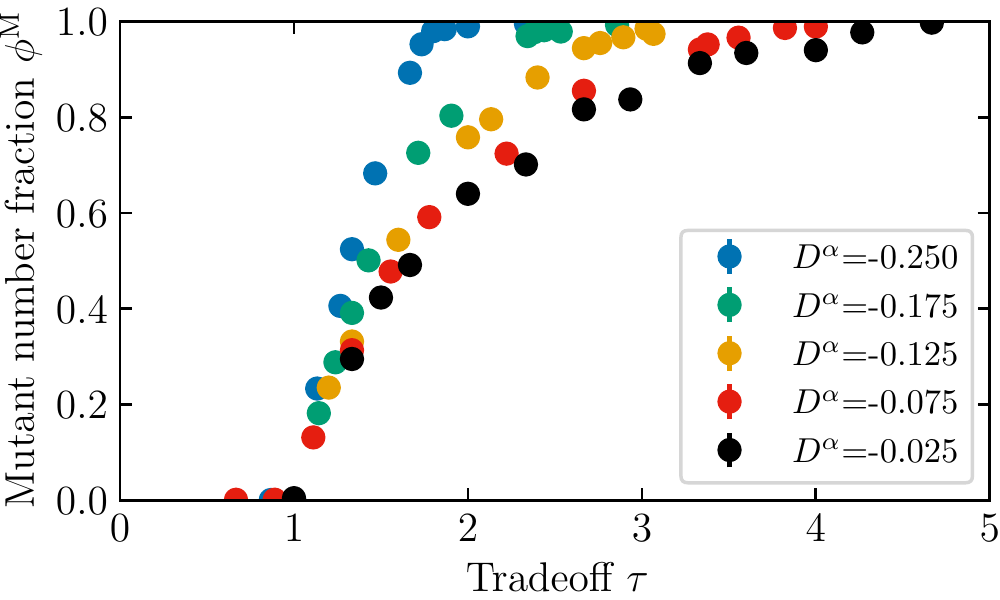}
		\includegraphics[width=0.49\linewidth]{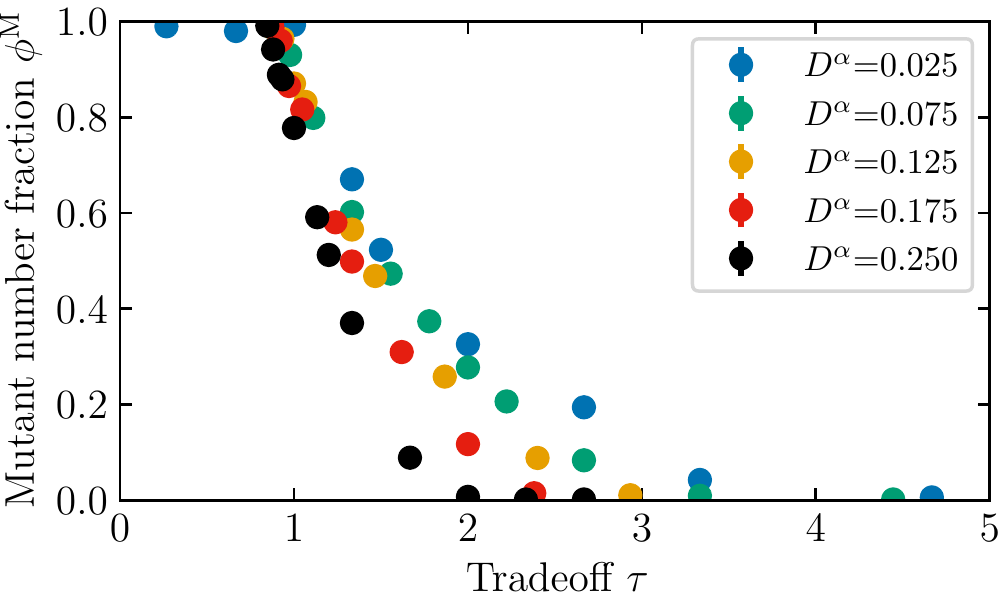}	
		
		\caption{a) Average number fractions of the mutant (same simulations as shown in Fig~\ref{number_fractions}a)) as a function of the tradeoff $\tau$ of Eq.~\eqref{genotype_values} for different evolutionary distances $D^\alpha$. b)~Same as in a) but with results from Fig~\ref{number_fractions}b). Error bars are obtained via block-averaging method (hidden behind markers).}
		\label{slope}
	\end{adjustwidth}
\end{figure} 

Figure \ref{slope} displays similar results as shown in Fig~\ref{number_fractions}, but now as a function of the tradeoff $\tau$ in Eq.~\eqref{genotype_values}. For $\tau<1$ the genotype with higher growth-force strength outcompetes the weaker genotype, for $1<\tau<2$ a transition towards the less adhesive genotype occurs, while for even higher values of the tradeoff $\tau >2$ the less adhesive genotype outcompetes the second genotype. This transition from strongly growing, adhesive to weakly growing, less adhesive genotypes is found in the same range of $\tau$ as in the competition between many genotypes. Hence, the simplified case of two competing genotypes captures the essential physics to explain the coexistence between many competing genotypes and, additionally, provides a quantitative description.
\begin{figure}
	\begin{adjustwidth}{-2.25in}{0in}
		\flushleft{\hspace{.2cm}(a)\hspace{9.cm}(b)}\\
		
		\includegraphics[width=0.49\linewidth]{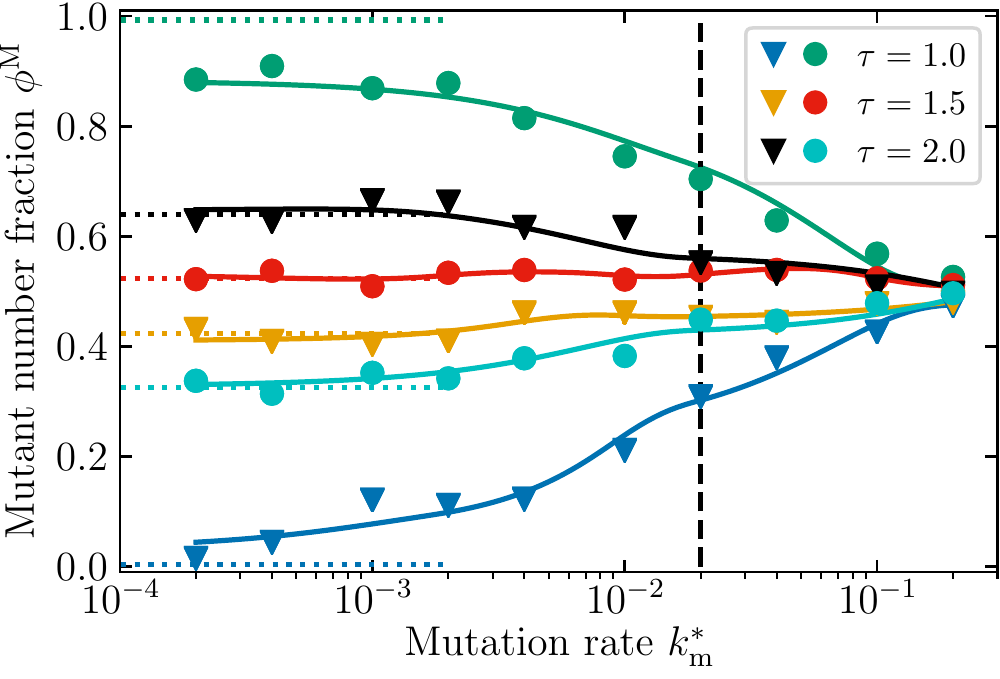}	
		\includegraphics[width=0.49\linewidth]{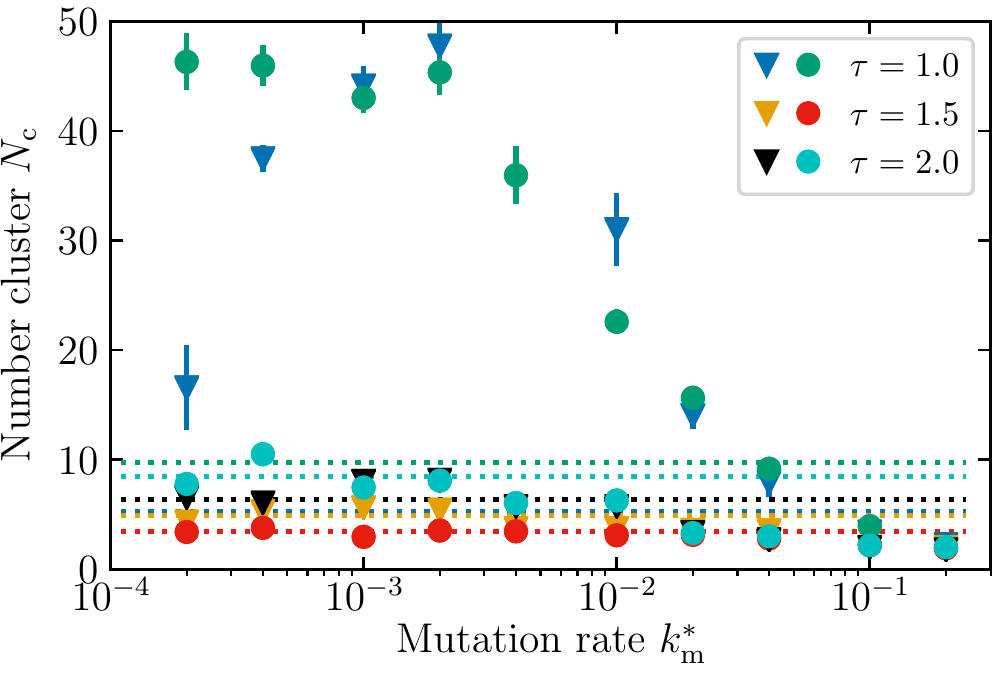}
		
		\caption{a) Average number fraction $\phi^{\text{M}}$ of the mutant as a function of the mutation rate $k^*_{\text{m}}$ for different values of the tradeoff $\tau$, for evolutionary distances $D^\alpha=0.025$ (circles) and $D^\alpha=-0.025$ (triangles). Horizontal dotted lines display results for a single mutation event. Vertical black dashed line indicates standard mutation rate. Solid lines are a guide to the eye. b)~Average number of clusters of the weaker genotype $N_{\text{c}}$ measured in the same competitions in a). Horizontal dotted lines display results for a single mutation event. Error bars are obtained via block-averaging method.}
		\label{mutationrate_cluster}
	\end{adjustwidth}
\end{figure}

Next, we turn our attention to the effect of a finite mutation rate on the evolution of the system.
Figure~\ref{mutationrate_cluster}a) shows the number fraction of the mutant as a function of $k_{\text{m}}$ for different combinations of evolutionary distance $D^\alpha$ and tradeoff $\tau$, in comparision to the number fraction reached for a single mutation event. As expexted, the number fraction converges towards $1/2$ with increasing $k_{\text{m}}$ for all combinations. For moderate mutation rates, however, the number fraction largely fluctuates around the same average as of a single mutation event. The single mutation leads to a stable coexistence of the two genotypes - additional mutations quickly relax back to this state. 
Siginificant deviations occur only if in the steady state of the single mutation event the number fraction of the weaker genotype is close to zero. In that case, the weaker genotype  consists only of one or very few small cohesive clusters of cells, because cells of the weaker genotype need to detach from the primary cluster in order to form new clusters, but are likely to die when they do so, as they are only surrounded by cells of the stronger genotype. Hence, the distribution of cells is highly non-homogenous. Compared to the single mutation event, even a small mutation rate leads to the formation of multiple small cluster all over the system, thus increasing the number fraction of the weaker genotype (see Fig~\ref{mutationrate_cluster}b) for comparision in terms of number of clusters and Materials and methods for further discussion).
This result explains why at least two genotypes, in addition to the dominating genotype, are populated as well in the cases shown in Fig~\ref{heterogeneity}a) and d). When the number fractions of both genotypes are sufficiently large (for $1\leq\tau\leq 2$), deviations from the average of a single mutation are still small for the standard mutation probability. Additionally, in the competitions between many genotypes, mutations change the genotype to $\alpha \pm 1$ randomly and not in a preferred direction. Hence, we conclude that the precise value of the mutation probability does not play an important role in the regime where we find a heterogeneous distribution of genotypes, as long as it is reasonably small ($k_{\text{m}} \ll k_{\text{a}}$).

\begin{figure}
	\begin{adjustwidth}{-2.25in}{0in}
	\flushleft{\hspace{.2cm}(a)\hspace{9.cm}(b)}\\
	\includegraphics[width=0.49\linewidth]{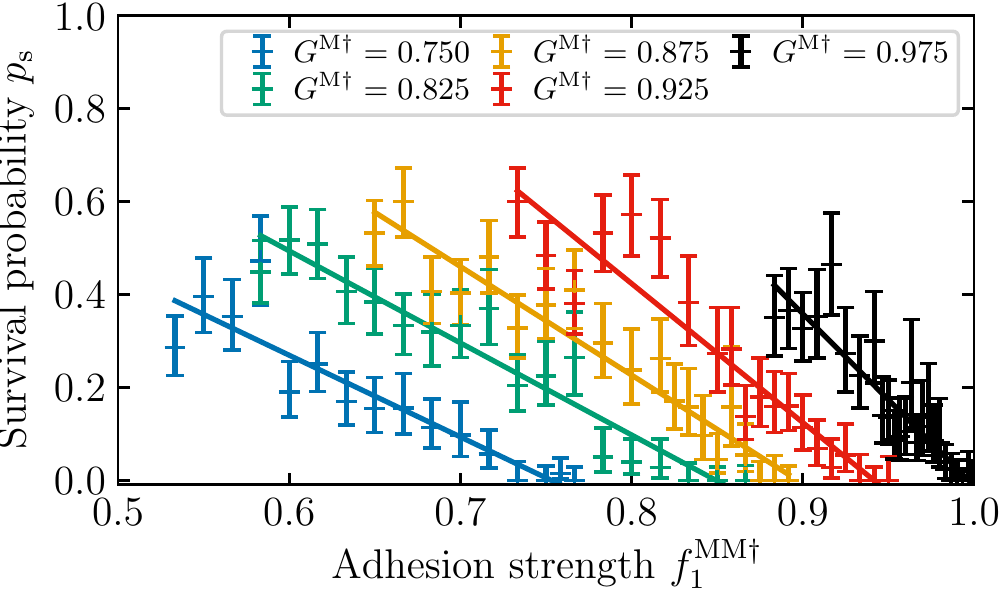}
	\includegraphics[width=0.49\linewidth]{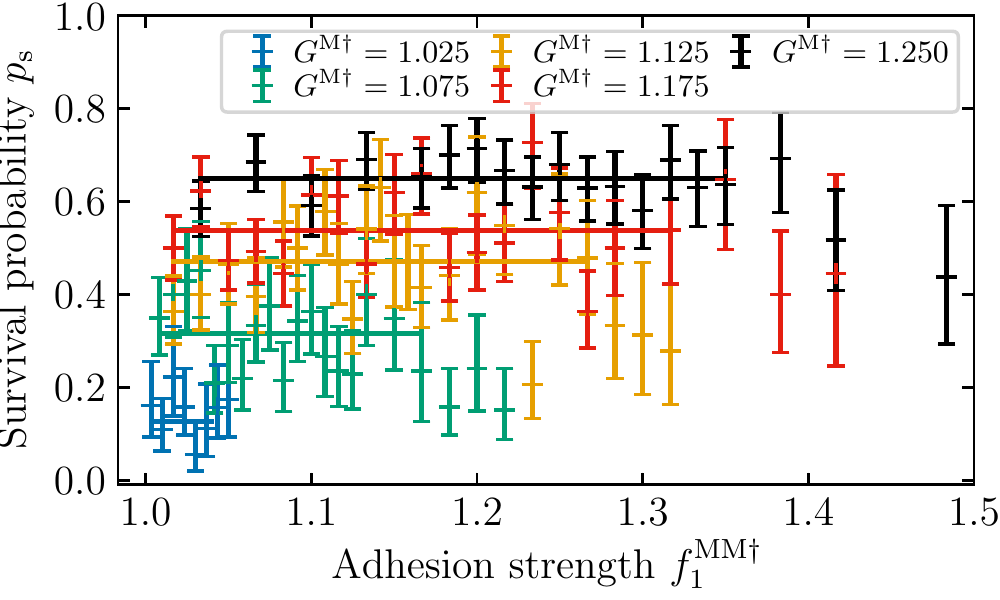}
	
	\caption{a) Probability $p_{\text{s}}$ of a single mutated cell to grow to macroscopic size as a function of its (reduced) adhesion strength $f_1^{\text{MM}\dagger}$ for several values of the (reduced) growth-force strength $G^{\text{M}\dagger}$. Error bars are a $1\sigma $ binomial confidence interval obtained by Clopper-Pearson. Solid lines are linear fits. b)~Same as in a), but with a constant fitted in the plateau regime and increased growth force and adhesion strength. Box size $L^*=8$.}
	\label{survival}
	\end{adjustwidth}
\end{figure}

Given that a single mutated cell can grow to tissue of macroscopic size in a certain parameter regime for $f_{1}^{\text{MW}}=\min(f_1^{\text{MM}},f_1^{\text{WW}})$, the question arises how likely it is to actually reach this state. In order to study this probability, we mutate again a single cell in a host tissue at its homeostatic state. A mutation that reaches a certain threshold $N_{\text{t}}=20$ of cells  counts as a survival event (the chance to die after reaching this treshold becomes extremely small), apoptosis of the last mutant cell as a death event. Figure~\ref{survival} shows the averages of many such simulations. For reduced growth force and adhesion strength, the survival probability $p_{\text{s}}$ is only non-zero below the critical adhesion strength $f_1^{\text{crit}}$. For $f_1^{\text{MM}}<f_1^{\text{crit}}$, $p_{\text{s}}$ increases linearly with further decreasing adhesion strength. On the other hand, when growth force and adhesion strength are increased, the survival probability first shows a plateau, whose value increases with increasing growth force strength, from which it will probably drop to zero with further increase. Simulations in this regime are difficult, because a mutated cell can easily grow to a few cells, but will hardly reach the number threshold nor completely vanish again. Due to the high self-adhesion strength on the one hand, it becomes hard to detach from the other cells, but on the other hand easy to grow against the host when only few or no other mutant cells are around. This explains the larger error bars at the highest values of the adhesion strength, where the sample size is small. 

\section*{Discussion}
 We have shown how intra-tumor heterogeneity, the existence of multiple subpopulations within the same tumor, can arise due to mechanical interactions alone. The simultaneous change of the adhesion and growth-force strength stabilizes the coexistence of multiple subpopulations, in a highly dynamic state.  A higher growth-force strength alone, as well as a lower adhesion strength, favor proliferation of a single subpopulation and the evolution of the system to cell types with the highest growth-force strength, or lowest adhesion strength, respectively. A tradeoff between the two, however, yields coexistence between multiple subpopulations of different cell types. Interestingly, the expression of the adhesion protein E-cadherin, which also affects cell growth, has been found to be down-regulated in many real tumors \cite{Beavon2000}.

The simulations also reveal that the homeostatic pressure of a cell type is not necessarily the only quantity that determines the result of a competition.
Interactions between different cell types, in our model determined by the adhesion between them, can lead to a completely reverse outcome, i.e. a cell type with lower homeostatic pressure can outcompete a stronger one completely.  A phenomenological model explains the results on a qualitative level. The evolution of each cell type is governed by mechanically-regulated growth, while mutation rates only play a minor role in the dynamics. 

An interesting future aspect to be studied is the influence of open boundaries. A tissue with a negative homeostatic pressure then naturally grows to a spheroid of finite size, with an enhanced rate of division at the surface \cite{Podewitz2015}. For competing cell types, this would lead to an interplay between surface and interfacial effects.
\section*{Materials and methods}
\subsection*{Standard (host) tissue and simulation parameters}
We define a set of reference simulation parameters, which we refer to as host parameters. Table \ref{standard} shows the values in simulation units. In simulations we keep the host W fixed and vary the parameters of the mutant M around the values of the host.
\begin{table}[h]
	\caption{Simulation parameters and measured properties of the standard (host) tissue}.	\label{standard}
	\raggedright
	\begin{tabular}{|l|l|l|}
		\hline
		Parameter & Symbol & Value \\ \thickhline
		Time Step & $\Delta t$ & $10^{-3}$\\ \hline
		Pair potential interaction range & $R_{\text{PP}}$ & $1$ \\ \hline
		Cellular expansion pressure constant & $r_0$ & 1 \\ \hline
		Cell division distance treshold & $r_{\text{ct}}$ & $0.8$ \\ \hline
		New cell particle initial distance & $r_{\text{d}}$  & 0.00001 \\ \hline
		Growth-force strength & $G$ & $40$ \\ \hline
		Mass & $m$ & 1 \\ \hline
		Intracell dissipation coefficient & $\gamma_{\text{c}}$ & $100$ \\ \hline
		Intercell dissipation coefficient & $\gamma_{\text{t}}$ & $50$ \\ \hline
		Background dissipation coefficient & $\gamma_{\text{b}}$ & $0.1$ \\ \hline
		Apoptosis rate & $k_{\text{a}}$ & 0.01 \\ \hline
		Mutation propability & $p_{\text{m}}$ & 0.01 \\ \hline
		Noise intensity & $k_{\text{B}}T$ & 0.1 \\ \hline
		Repulsive cell-cell potential coefficient & $f_0$ & $2.39566$ \\ \hline
		Attractive cell-cell potential coefficient & $f_1$ & $6.0$ \\ \hline
		Isothermal compressibility & $\beta_{\text{T}}$ & 1 \\ \hline
		Relaxation time constant & $t_{\text{P}}$ & $1$ \\ \hline
		Homeostatic pressure & $P_{\text{H}}^*$ & $\SI{0.1321 \pm 0.0005}{}$ \\ \hline
		Pressure response coefficient &$\kappa^*$ &  $\SI{2.676\pm 0.080}{}$  \\ \hline
	\end{tabular}
	\label{standard}
\end{table}

\subsection*{Cluster analysis}
As explained in the results section, a constant rate of mutation leads to an enhanced formation of clusters when the weaker genotype is barely able to grow against the stronger genotype and consists of only one or few clusters for a single mutation event. 
We define a cluster as all cells of the same genotype that are in interaction range to at least one other member of the cluster (DBSCAN clustering algorithm with number of minimal points equal to one). Figure \ref{mutationrate_cluster}b) displays the number of clusters of the weaker genotype in the competitions displayed in Fig~\ref{mutationrate_cluster}a), in comparison to the result of a single mutation event. Indeed, when the number fraction of the weaker genotype is small for a the single mutation event ($\tau=1$), we find significant deviations even for small mutation rates. In this case, the number of clusters first strongly increases with mutation rate, with roughly a tenfold increase at the peak. For even higher mutation probability, the number of clusters decreases again, due to merging of clusters, finally leading to percolation.

\section*{Supporting information}

\paragraph*{S1 Appendix.}
\label{S1_Appendix}
{\bf Additional Results.} Appendix containing additional results and the corresponding figures.



\section*{Acknowledgments}
The authors gratefully acknowledge the computing time granted through JARA-HPC on the supercomputer JURECA \cite{jureca} at Forschungszentrum Jülich.


%
%
%

\end{document}